\documentclass[%
reprint,
amsmath,amssymb,
aps,
pra,
nofootinbib
]{revtex4-2}

\usepackage{graphicx}
\usepackage{dcolumn}
\usepackage{bm}
\usepackage{lipsum}

\usepackage[utf8]{inputenc}
\usepackage[T1]{fontenc} 
\usepackage{mathtools}
\usepackage{amsmath}
\usepackage{amssymb}
\usepackage{amsthm}
\usepackage{dsfont}
\usepackage{graphicx,graphics,caption,subcaption}
\usepackage{lipsum}
\usepackage{esvect}
\usepackage{float}
\usepackage{bbold}
\usepackage{appendix}
\usepackage{xcolor}
\usepackage{hyperref}

\usepackage{graphicx,graphics,caption,subcaption}
\usepackage{multirow}
\usepackage{diagbox}

\hypersetup{
	colorlinks=true,                
	breaklinks=true,                
	urlcolor= black,                
	linkcolor= blue,                
	bookmarksopen=false,
	filecolor=black,
	citecolor=black,
	linkbordercolor=blue,
	citebordercolor={0 1 0}
	}

\AtBeginDocument{\hypersetup{pdfborder={0 0 1}}}
\hypersetup{pdftitle = {Upper bound on the Guessing probability using Machine Learning},
       pdfauthor = {Sarnava Datta, Hermann Kampermann, Dagmar Bruß},
       pdfsubject = {Quantum information theory}, 
       pdfkeywords = {quantum, theory, 
                      SDP, semidefinite program, linear program, nonlocality, Bell,
                      entanglement, device-independent quantum key distribution, quantum correlation, guessing probability, facet, classical polytope, Bell scenario, DIQKD, device-independent randomness generation, DIRNG, machine learning, ML, deep learning, DL, feed forward neural network, supervised machine learning, neural network, weighted vertex sampling, upper bound, Bell inequality, 
                      convex, optimization, dual,
                      , 
              }
      }
\begin{document}
\preprint{APS/123-QED}

\title{Upper bound on the Guessing probability using Machine Learning} 
\author{Sarnava Datta}
\email{Sarnava.Datta@hhu.de}
\author{Hermann Kampermann}
\author{Dagmar Bru\ss}
\affiliation{ Institut f\"{u}r Theoretische Physik III\\
    Heinrich-Heine-Universit\"at D\"usseldorf}
\date{\today}

\begin{abstract}
The estimation of the guessing probability has paramount importance in quantum cryptographic processes. It can also be used as a witness for nonlocal correlations. In most of the studied scenarios, estimating the guessing probability amounts to solving a semi-definite programme, for which potent algorithms exist. However, the size of those programs grows exponentially with the system size, becoming infeasible even for small numbers of inputs and outputs. We have implemented deep learning approaches for some relevant Bell scenarios to confront this problem. Our results show the capabilities of machine learning for estimating the guessing probability and for understanding nonlocality.
\end{abstract}	
\maketitle 

\section{Introduction}	
Whenever the statistics of a measurement on a composite quantum state contradict the assumptions of local realism, thus violating a Bell-type inequality, the correlations are referred to as nonlocal \cite{bell1964einstein}.  
These nonlocal correlations are used to certify private randomness in  device-independent quantum key distribution (DIQKD) \cite{acin2007device,pironio2009device,arnon2019simple,arnon2018practical,barrett2005no,masanes2011secure,vazirani2014fully,masanes2014full,acin2006efficient,murta2019towards,holz2020genuine,hanggi2010device} and device-independent randomness generation (DIRNG) \cite{pironio2010random,nieto2018device,pironio2013security,bancal2014more,nieto2014using,bischof2017measurement,acin2012randomness,acin2016certified,skrzypczyk2018maximal}. For quantifying randomness, estimating the guessing probability is often an important task. The guessing probability is the probability with which an adversary can guess an outcome of another party's measurement. 
If the guessing probability is less than 1, the adversary cannot predict the outcome with certainty. This implies the presence of intrinsic randomness in the system. However, bounding the guessing probability is not an easy task. Typically it is not possible to explicitly compute the guessing probability, but one can only provide an upper bound by solving a semi-definite optimization problem. Usually, one bounds the guessing probability from a given Bell inequality, and the corresponding quantum violation  \cite{masanes2011secure,pironio2010random}. Here, one needs to use the hierarchical structure of the quantum correlations \cite{navascues2007bounding,navascues2008convergent} to solve the semi-definite optimization problem. 
The complexity of this optimization problem is increasing and becoming computationally demanding with the number of settings and outcomes.  

In this paper, motivated by the outstanding recent progress in utilizing machine learning in the field of quantum information \cite{carrasquilla2017machine, broecker2017machine, canabarro2019machine, carleo2017solving, deng2018machine, gao2017efficient, ma2018transforming, mehta2019high, torlai2018neural, venderley2018machine}, we develop deep learning (DL) models that predict the guessing probability along with the optimal Bell inequalities (used to upper bound the guessing probability) from an observed probability distribution using supervised machine learning. 
A crucial element of supervised machine learning is to generate sample data input and output to train the model. Here, we sample random quantum probability distributions and use them as the input of the training data. With this data, using the two-step method of Ref. \cite{datta2021device}, we estimate the upper bound of the guessing probability and the optimal Bell inequality, and use it as the output of the training data. After sufficient training, our DL approach can recognize the pattern and predict the guessing probability and the optimal Bell inequality with high accuracy and low average statistical error.

We organize this work as follows. We start in Sec.~\ref{Bell setup} by explaining the generalized Bell set-up, types of correlations and Bell inequalities. We introduce the guessing probability and show how to estimate it by solving a semi-definite programme in Sec.~\ref{Guessing probability}. We introduce our deep learning approach in Sec.~\ref{sec ML approach}. We discuss how to sample quantum probability distributions from the quantum correlation space, which are then used as input for supervised learning. We build several deep learning models for predicting the guessing probability and the Bell inequality for various Bell scenarios and measure their efficiency to show the model's utility. 
	
\section{Generalized Bell Set-up} \label{Bell setup}
In this section, we introduce a generalized Bell set-up. In each measurement round, two parties, Alice and Bob, share a quantum state $ \rho_{AB} $ acting on $ \mathcal{H}_{A} \otimes \mathcal{H}_{B}$. In the presence of an eavesdropper Eve, her side information $ E $ is described via the purification of the joint system $ \rho_{ABE} $ acting on $ \mathcal{H}_{A} \otimes \mathcal{H}_{B} \otimes \mathcal{H}_{E}$ where $ \mathrm{Tr}_{E}\left(\rho_{ABE}\right)=\rho_{AB} $. Each party selects locally an input (a measurement setting) which produces an output (a measurement outcome). We refer to this scenario as a Bell scenario.
\begin{figure}[htb] 
\includegraphics[height=7cm,width=9cm]{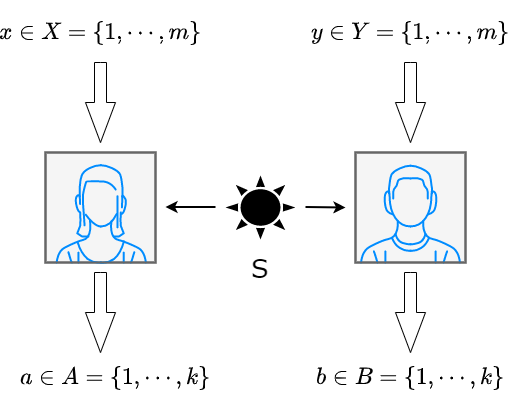}
\caption{Schematic description of a Bell scenario consisting of two parties, Alice and Bob. For further explanation, see the main text. 
}
\label{Bell Scenario figure}
\end{figure}
Alice performs measurements specified by her input $x \in X = \{1,\cdots,m\}$, where each input has $k$ possible outcomes $a \in A= \{ 1, \cdots, k \}$. Similarly, Bob performs measurements specified by his input $y \in Y = \{1,\cdots,m\}$ and produces the outputs $ b \in B= \{ 1, \cdots, k \}$. We denote this scenario as $[m,k]$ Bell scenario, i.e. $ m $ measurement settings with $ k $ outcomes each; see Fig.~\ref{Bell Scenario figure} for visualization. After many repetitions, the conditional probability $P(ab|xy)$ can be estimated. The Bell scenario is completely characterized by the set $\textbf{P}:=\{ P(ab|xy) \} \subset \mathds{R}^{m^2k^{2}}$ of all joint conditional probabilities which we refer to as a behavior \cite{brunner2014bell}. Thus, the following constraints are imposed: positivity $P(ab|xy) \geq 0$ $\forall$ $a,b,x,y$ and the normalization $ \sum_{a,b=1}^{k} P(ab \arrowvert xy)=1$ for all $x$ and $y$. We say the behavior is no-signaling if the input-output correlation obeys
\begin{equation}\label{No-signalling eq}
\begin{aligned}
    \sum_{b=1}^{k}P(ab|xy) &= P(a|x) \qquad \forall a,x,y \quad \textup{and} \\
    \sum_{a=1}^{k}P(ab|xy) &= P(b|y) \qquad \forall b,x,y \, .
\end{aligned}    
\end{equation}
The set of all correlations satisfying the no-signaling constraints forms a convex polytope $\mathcal{NS}$. A behavior is said to be local if it can be written as a convex mixture of deterministic strategies \cite{pitowsky1982resolution,pitowsky1991correlation}. The set of all local correlations forms a convex polytope $\mathcal{P}$. There exist inequalities of the form \cite{brunner2014bell} 
\begin{equation}\label{Bell Inequality}
  \sum_{a,b,x,y} C_{abxy}P(ab|xy) \leq \mathcal{I}_{L} \, ,
\end{equation}	
which separate the set of all local correlations (in other words, the convex polytope $\mathcal{P}$) from the nonlocal behaviors. These inequalities are called Bell inequalities. A Bell inequality is specified by the coefficients $C_{abxy} \in \mathds{R}$. We denote a Bell inequality as $B$, and $ \sum_{a,b,x,y}C_{abxy}P(ab|xy) $ as the Bell value $ B[\textbf{P}] $ in this paper. Here, $\mathcal{I}_{L}$ is the classical bound, which is the maximal value over all local behaviors. Thus, a behavior with a classical origin, i.e. $ \{P(ab|xy)\} \in \mathcal{P}$, cannot violate this inequality. 

The Born rule of quantum theory postulates that a behavior is quantum if there exists a quantum state $\rho_{AB}$ acting on a joint Hilbert space $ \mathcal{H}_A \otimes \mathcal{H}_B$ of arbitrary dimension and measurement operators (POVM elements) $\{M_{a|x}\}$ with $ M_{a|x} \ge 0 $ and $ \sum_a M_{a|x} = \mathds{1} $ $\forall x$, and $\{M_{b|y}\}$ with analogous properties such that 
\begin{equation}
P(ab|xy) = \mathrm{Tr}(\rho_{AB} M_{a|x}\otimes M_{b|y})\, .
\end{equation}
The set of all quantum correlations forms a convex set $\mathcal{Q}$. If a behavior $ \{P(ab|xy)\} \in \mathcal{Q} \setminus \mathcal{P}$, it violates at least one Bell inequality of the form in Eq.~(\ref{Bell Inequality}). 
\begin{figure}[h]
\centering
\includegraphics[height=9cm, width=10cm]{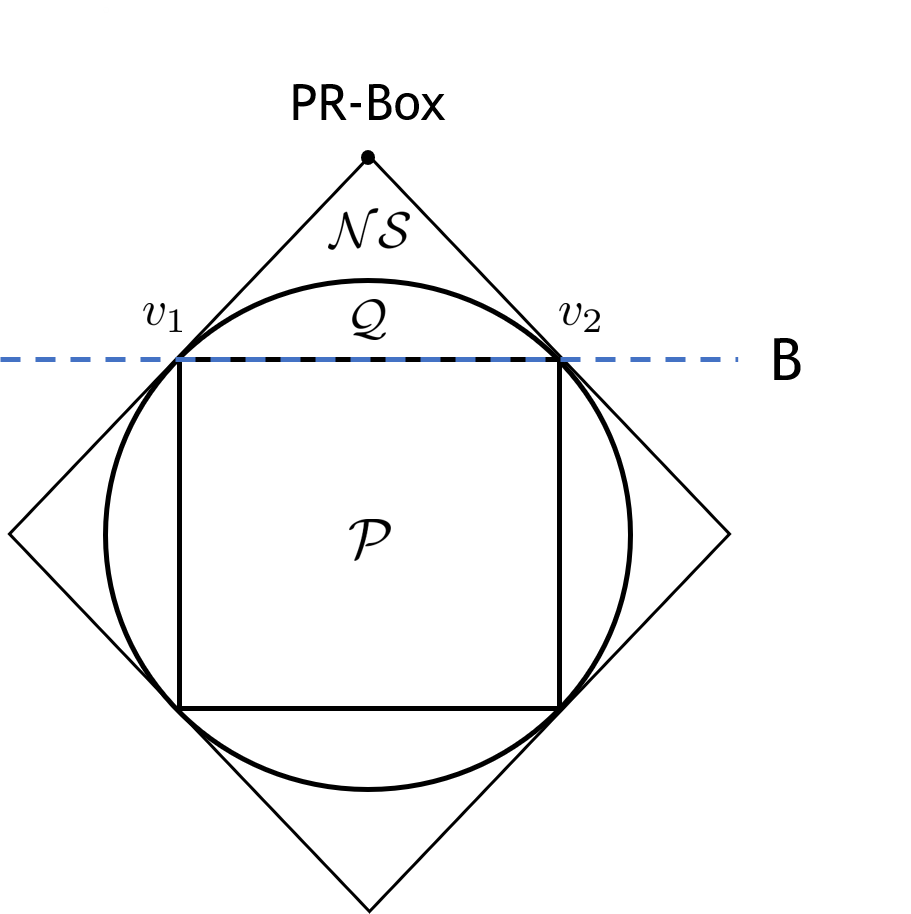}
\caption{A pictorial representation for the set of correlations. All classical probabilities form a convex polytope $\mathcal{P}$, which is embedded in the set $\mathcal{Q}$ of quantum correlations, which in turn is a subset of the no-signaling polytope $\mathcal{NS}$. $v_1$ and $v_2$ are vertices of the local polytope. $B$ (blue dashed line) represents the Bell inequality which separates the classical polytope from the quantum and no-signaling set.}
\label{CQNS}
\end{figure}
The sets $ \mathcal{P} $, $ \mathcal{Q} $ and $ \mathcal{NS} $ obey the following relation: $ \mathcal{P}  \subsetneq \mathcal{Q} \subsetneq \mathcal{NS} $; see Fig.~\ref{CQNS} for a pictorial representation.

\section{Guessing probability} \label{Guessing probability}
In an adversarial black box scenario framework, the adversary Eve tries to guess some outcomes obtained by Alice and Bob. The probability that Eve can correctly guess the outcome is called the guessing probability. Here, we denote the guessing probability as $ P_g(a|x,E) $, which is the guessing probability of Eve about Alice's outcome $a$ corresponding to her measurement setting $x$. In Ref.\cite{masanes2011secure}, it is shown that $ P_g(a|x,E) $ can be upper bounded by a function $ G_x $ of the observed Bell value $B[\textbf{P}]$ of a particular Bell inequality $B$ by semi-definite programming, i.e. $P_g(a|x,E) \leq G_x(B[\textbf{P}])$. One crucial element to bound the guessing probability $ P_g(a|x,E) $ is to choose a suitable Bell inequality. We follow the two-step procedure of \cite{datta2021device} where the Bell inequality is constructed from the input-output probability distribution \textbf{P} that leads to the maximum Bell violation for that particular measurement statistics. 

This is achieved by solving the linear program:
\begin{equation}\label{linear_optimization}
\begin{aligned}
    & \underset{\textbf{h},c}{\text{max}}
    & & \textbf{h}^T \textbf{P}-c \, ,\\
    & \text{subject to}
    & & \textbf{\textbf{h}}^T \textbf{v}_{p} \leq c \quad \forall \quad p \in \{1,\cdots,k^{2m} \} \, , \\
    &&& \textbf{h}^T \textbf{P} > c  \, ,\\
    &&& -1 \leq h_i \leq 1 \quad \forall \quad i \in \{1,\cdots,m^2k^2 \}\, .
\end{aligned}
\end{equation}
Here $\textbf{h}$ is the hyperplane specifying the Bell inequality $B$, \textbf{P} denotes the measurement data, $\textbf{v}_{p}$ corresponds to the $p$ vertices of the classical polytope $\mathcal{P}$ and $c$ is the classical bound. Thus the Bell inequality $B$ found by the optimization of Eq.~(\ref{linear_optimization}) and specified by the hyperplane vector \textbf{h} is given as:
\begin{equation}\label{Hyperplane BI}
\sum_{a, b, x, y} h_{abxy}P(ab|xy) \leq c \, ,
\end{equation}
where $a \in A$, $b \in B$, $x \in X$, $y \in Y$. We will use the Bell inequality $B$ and corresponding Bell value $B[\textbf{P}] =\sum_{a, b, x, y} h_{abxy}P(ab|xy)$ to upper bound the guessing probability $P_g(a|x,E)$ by solving the following semidefinite program \cite{masanes2011secure}:
\begin{equation}\label{SDP}
\begin{aligned}[t]
    & \underset{ \rho_{AB},\{A(a|x)\}, \{B(b|y)\}}{\text{max}}\quad P_{g}(a|x,E)\\
    & \text{subject to:} \quad
    \mathrm{Tr}(\rho_{AB} \mathcal{G})=B[\textbf{P}] \, .
\end{aligned}  
\end{equation}
In the optimization problem of Eq.~(\ref{SDP}), the guessing probability is bounded using the NPA-hierarchy \cite{navascues2007bounding,navascues2008convergent} up to level 2. The optimization is performed using standard tools YALMIP \cite{Lofberg2004}, CVX \cite{gb08,cvx} and QETLAB \cite{qetlab}.
Note that, $A(a|x)$ and $B(b|y)$ are the measurement operators of Alice and Bob, respectively, and $\rho_{AB}$ is the state shared between them. $\mathcal{G}$ is the Bell operator defined as:
\begin{equation}
	\mathcal{G}=\sum_{a, b, x, y} h_{abxy} A(a|x) B(b|y) \, .
\end{equation}
Let us denote $P^{*}_{g}(a|x,E)$ as the upper bound of the guessing probability, which is the solution to the optimization problem of Eq.~(\ref{SDP}).

\section{Machine learning approach} \label{sec ML approach}
Providing an upper bound for the guessing probability by solving a semi-definite program is a computationally demanding task. It becomes arduous when the Bell scenario raises its complexity, i.e. for an increased number of measurement inputs and/or outputs in the Bell scenario.

Thus, in this paper, we approach solving the problem via machine learning (ML) (see Ref. \cite{Goodfellow-et-al-2016} for detailed discussions on the concepts of machine learning) so that the trained model can estimate the guessing probability $P^{*}_g(a|x,E)$ from the input-output probability distribution $\{P(ab|xy)\}$. We are going to use the supervised learning technique. In a supervised ML approach, the first step is generating the training points. We use random bipartite quantum probability distributions as the supervised ML model's input (features), after generating them from facet Bell inequalities using the \textit{weighted vertex sampling} method \cite{krivachy2021high}. Since the guessing probability for local behaviors is always 1 (i.e. Eve can guess the right outcome with probability 1), we do not need to train the machine to perform well on those. Thus we only take samples from the nonlocal part of the no-signaling set, i.e. $\mathcal{NS} \setminus \mathcal{P}$. 
To single out the input-output correlation with a quantum realization, we reduce the samples using the NPA hierarchy to approximate the quantum realizable probability distribution.

Explicitly, we generate samples from the quantum set $\mathcal{Q}$ as follows.
For the $[m,k]$ Bell scenario (i.e. $m$ measurements, $k$ outcomes each), the classical polytope $\mathcal{P}$ is specified by $k^{2m}$ local vertices.
The classical polytope can also be described by its facets, which represent the hyperplanes (or Bell inequalities) that separate any non-classical (quantum and no-signaling) behavior from the classical ones. These facets are called facet Bell inequalities or tight Bell inequalities \cite{brunner2014bell}; see Fig.~\ref{CQNS} for a pictorial representation.  
For the $[2,2]$ scenario, eight facet Bell inequalities exist, all equivalent to the CHSH inequality \cite{clauser1969proposed}. 
For the $[3,2]$ Bell scenario, there are 648 facet Bell inequalities. These facet Bell inequalities are found using the formulation of Ref. \cite{fukuda2003cddlib} \footnote{Using Ref. \cite{fukuda2003cddlib}, one can calculate all the facets of a convex polytope given its vertices. The transformation from the vertex representation to the facet representation of a polytope is known as facet enumeration or convex hull problem, which uses Gaussian and Fourier-Motzkin elimination. The list of facets consists of positivity constraints and the facet Bell inequalities. 
Here, we only focus on the facet Bell inequalities alone.}. Note that all the 648 facet Bell inequalities correspond to two classes of independent facet Bell inequalities, i.e. the CHSH inequalities and the $I3322$ inequalities \cite{collins2004relevant,pal2010maximal}. We consider all facet Bell inequalities for the [2,2] and [3,2] Bell scenario while generating training points for the supervised machine learning problem.
For the $[4,2]$ Bell scenario, there are 174 independent facet Bell inequalities \cite{cruzeiro2019complete}. Since there are many (>10000) equivalent facets  \cite{cope2019bell}, we will only consider the independent ones. 
These facet Bell inequalities are spanned by some of the local vertices of the classical polytope \footnote{For the $[2,2]$ Bell scenario, all the facet inequalities are spanned by eight local vertices. For the $[3,2]$ Bell scenario, facet Bell inequalities, equivalent to the CHSH inequality, are spanned by thirty-two vertices. Twenty vertices span the inequalities equivalent to the $I3322$ inequality.}. These vertices provide the maximum classical bound of the corresponding facet Bell inequality. Consider the case that $n$ local vertices span a facet Bell inequality, where we denote the set of $n$ vertices as $\left\{P_{i}^{\mathcal{L}}(ab|x y)\right\}_{i=1}^{n}$. We denote the PR-box of the corresponding facet Bell inequality as $P^{\mathrm{PR}}(ab|xy)$, see Fig.~\ref{CQNS} for visualization. The PR-Box $P^{\text{PR}}(ab|xy)$ can be defined as the probability distribution that provides the maximal no-signaling bound of the corresponding facet Bell inequality \cite{barrett2005nonlocal,popescu1994quantum}. We take uniformly random weighted mixtures of the $n+1$ vertices ($n$ vertices that span the facet Bell inequality and the corresponding PR-box) with an $n$-fold weight on the PR-box.
Formally, the sample behavior from the set $\mathcal{NS}\setminus \mathcal{P}$ can be generated as:
\begin{equation}
\textbf{P}:=P(a b|x y)=\frac{n w_{0} P^{\mathrm{PR}}(ab|xy)+\sum_{i=1}^{n} w_{i} P_{i}^{\mathcal{L}}(ab|xy)}{n w_{0}+\sum_{i=1}^{n} w_{i}}
\end{equation} 
where the $w_{i} \in [0,1]$ with $i=0,1,\cdots,n$, are uniformly drawn random numbers. This process is done for all facet inequalities.
From this set of samples, we only select the ones with a $\mathcal{Q}_2$ realization (the second level of NPA hierarchy \cite{navascues2007bounding,navascues2008convergent}). Here we work under the assumption that $\mathcal{Q}_2$ provides a good approximation for the original quantum set $\mathcal{Q}$.

\begin{figure}[H]
\centering
\includegraphics[height=8cm, width=4cm]{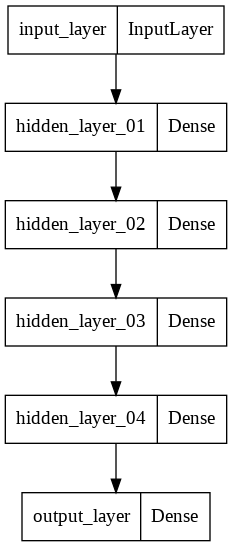}
\caption{Schematic description of a 'linear' neural network. It consists of an input layer, several hidden layers and an output layer without branching. The hidden layers and the output layer are dense layers, meaning that the neurons of the layer are connected to every neuron of its preceding layer.}
\label{FFNN_oneNN}
\end{figure}

We store the probability distribution \{$ P(ab|xy) $\} and use it as the input (features) of the supervised machine learning problem, i.e.
\begin{equation}\label{input of model}
	\textbf{X}:=\{ P(ab \arrowvert xy) \}^{a,b=1,\cdots,k}_{x,y=1,\cdots,m}  \, .
\end{equation}
We calculate the guessing probability of each input \textbf{P} using the two-step method (see Sec.~\ref{Guessing probability} for details),
and use it as the output (target), i.e.
\begin{equation} \label{output of model 1}
	y= P^{*}_g(a|x,E)\, .
\end{equation}
Without loss of generality, we have always calculated the guessing probability of Alice's first measurement setting. We use a deep neural network to assess the dataset and make predictions. We fed the input-output pair $\{\textbf{X},y\}$ (see Eq.~(\ref{input of model}) and Eq.~(\ref{output of model 1})) into an artificial neural network (ANN) to learn the best possible fit. For an elaborate explanation of an artificial neural network, see Ref. \cite{Goodfellow-et-al-2016}. Following the standard approach, we divide the dataset into two parts. The first part of the dataset is for training and validation (80\%), and the second is for testing (20\%). We choose a 'linear' \footnote{Here, linear means that there is no branching in the hidden layers of the neural network architecture.} ANN with several layers as our model; see Fig.~\ref{FFNN_oneNN} for visualization.
The input layer has $m^2k^2$ neurons corresponding to the elements in \{$ P(ab\arrowvert xy) $\}. The output (last) layer has only one neuron since we only have to predict one element: the guessing probability $P^{*}_g(a|x,E)$. We perform 100 rounds of training using the optimizer ADAM \cite{kingma2014adam}, of which the first 50 rounds have a fixed learning rate of 0.001. For the next 50 rounds, we reduce the learning rate by 90\% in every tenth round. We choose the activation function ReLu (Rectified linear unit) \footnote{Relu activation function: $\phi(x)=max(0,x)$} in the input and the hidden layers while using the sigmoid activation function\footnote{sigmoid activation function: $\phi(x)=\frac{1}{1+e^{-x}}$} in the output layer. The ReLu activation function introduces non-linearity and the sigmoid activation function keeps the output within the range of [0,1].
The mean squared error (MSE) \footnote{$\text{MSE}(y,\hat{y})=\frac{1}{N}\sum_{i=0}^{N}\left(y_i-\hat{y}_{i}\right)^2$, where $y$ is the original output and $\hat{y}$ is the estimated output of the model.} is used as our loss function, which is minimized during the training process.
\begin{table}[h]
\centering
\begin{tabular}{|c|c|c| c|}\hline
    \diagbox[width=10em]{Metrics}{Bell \\ Scenario}&
    [2,2] & [3,2] & [4,2] \\ \hline
    MSE & 0.00007 & 0.007 & 0.009 \\ \hline
    MAE & 0.0003 & 0.01 & 0.025 \\ \hline
\end{tabular}
\caption{Performance measures for different Bell scenarios when estimating $P^{*}_g(a|x,E)$ from the probability distribution $P(ab \arrowvert xy)$. MAE: Mean Absolute Error (see Eq.~(\ref{Eq. MAE})), MSE: Mean Squared Error (see Eq.~(\ref{Eq. MSE}))}
\label{Table_prob_pGuess_error}
\end{table}
The trained model generates the predicted value of the guessing probability $ P^{\text{pred}}_g(a|x,E) $. To check the efficiency of our approach, we have used the mean absolute error (MAE)

\begin{equation}\label{Eq. MAE}
\begin{aligned}
&\text{MAE}\left[ P^{*}_g(a|x,E),P^{\text{pred}}_g(a|x,E) \right] \\
&= \frac{1}{N_{\mathrm{test}}}\sum_{i=1}^{N_{\mathrm{test}}} \left | P^{*}_g(a|x,E)_i-P^{\text{pred}}_g(a|x,E)_i \right| \, ,
\end{aligned} 
\end{equation}   
and the mean squared error (MSE)
\begin{equation}\label{Eq. MSE}
\begin{aligned}
&\text{MSE}\left[P^{*}_g(a|x,E),P^{\text{pred}}_g(a|x,E)\right] \\
&= \frac{1}{N_{\mathrm{test}}}\sum_{i=1}^{N_{\mathrm{test}}} \left( P^{*}_g(a|x,E)_i-P^{\text{pred}}_g(a|x,E)_i \right)^2 \, ,
\end{aligned}
\end{equation}
as a performance measure. $ N_{\mathrm{test}} $ is the number of data points in the test set. We analyze the results for different bipartite Bell scenarios and list the errors in Table~\ref{Table_prob_pGuess_error}. The average error is in the order of $ 10^{-4} $ to $10^{-2}$. Such high accuracy and small error without knowing the Bell inequality are truly remarkable.
\begin{table}[h]
\centering
\begin{tabular}{|c|c|c|c|}\hline
    \diagbox[width=10em]{Time\\ per sample}{Bell \\ Scenario}&[2,2] & [3,2] & [4,2] \\ \hline
    Mosek & 95ms & 496ms & 1568 ms \\ \hline
    Neural Network & 27$\mu$s & 35$\mu$s & 49$\mu$s \\ \hline
\end{tabular}
\caption{Runtime per sample comparison for SDP solver Mosek and the neural network method for estimating the guessing probability for different Bell scenarios.}
\label{runtime comparison 1}
\end{table}
We also compare the runtime performance of the neural network model with the frequently used SDP solver Mosek \cite{aps2019mosek} (that can be used to upper bound the guessing probability by solving the optimization problem of Eq.~(\ref{SDP})) in Table~\ref{runtime comparison 1}. The Mosek task is generated and solved using the Ncpol2sdpa \cite{wittek2015algorithm}. The results are evaluated over 10000 unknown samples and performed on a personal computer \footnote{Specifications: Intel(R) Core(TM) i7-10510U Processor, 2.30GHz  Frequency, 16.0 GB RAM} under comparable conditions. Once the neural network is trained, 
we get a speed-up of $10^{3}-10^{5}$ for obtaining a prediction about a new instance, compared to the runtime of the usual method for solving the optimization problem; see Table~\ref{runtime comparison 1}. 
This follows from the fact that the number of variables in the optimization process of Eq.~(\ref{SDP}) increases exponentially with the number of measurement settings (or outcomes per measurement) in the Bell scenario. Thus, it takes more computational time to perform the SDP using a classical solver like Mosek. A trained neural network only calculates the functional output using the optimized weights and biases. Only the neural network size affects the computational time needed to complete the prediction task.\\

However, the upper bound on the guessing probability calculated from a trained machine learning model only provides an estimate of its real value. Thus, we cannot use this estimate to bound the secret key rate. The predicted Bell inequality on the other side that generates a non-zero Bell violation (for a particular measurement statistics) can be used to bound the guessing probability (see Sec.~\ref{Guessing probability} for details) and the secret key rate. 
That's why in the next step, we use deep learning to predict the associated optimal Bell inequality $B$, which is then used to upper bound the guessing probability (see Sec.~\ref{Guessing probability} for details). For this purpose, we again use the neural network architecture where supervised learning is incorporated. We start by preparing the dataset where our input features are
\begin{equation}\label{input of model 2}
\textbf{X}:=\{ P(ab \arrowvert xy)\}^{a,b=1,\cdots,k}_{x,y=1,\cdots,m}  \, .
\end{equation}
Note that, the input is identical to Eq.~(\ref{input of model}).
The outputs are now the coefficients of the optimal Bell inequality $B$ (specified by $\{h_{abxy}\}^{a,b=1,\cdots,k}_{x,y=1,\cdots,m}$, see Eq.~(\ref{linear_optimization})) and the guessing probability $P^{*}_g(a|x,E)$, i.e. 
\begin{equation}\label{output of model 2}
\textbf{y}:=\left[\{h_{abxy}\}^{a,b=1,\cdots,k}_{x,y=1,\cdots,m},P^{*}_g(a|x,E)\right]\, .
\end{equation}
Here we use two types of neural network architectures. The first neural network is a usual 'linear' feed forward neural network (see \cite{Goodfellow-et-al-2016} for details, schematically represented in Fig.~\ref{FFNN_oneNN}). For $[m,k]$ Bell scenario, the input layer has $m^2k^2$ neurons (corresponds to the elements of $\{P(ab\arrowvert xy)\}$). The input layer is followed by several hidden layers. Unlike in the previous scenario, the output layer has $m^2k^2+1$ neurons in this case, where $m^2k^2$ neurons correspond to the coefficients of the Bell inequality $h_{abxy}$, and one neuron corresponds to the guessing probability $P^{*}_g(a|x,E)$. In this paper, we denote this construction of the 'linear' deep neural network as $\mathrm{NN}_1$. Following the standard approach, we divide the dataset $\{\textbf{X},\textbf{y}\}$ (see Eq.~(\ref{input of model 2}) and Eq.~(\ref{output of model 2})) into two sets; the first part of the dataset is for training and validation (80\%), and the second part is for testing (20\%). Similar to the training of the previous network, we perform 100 rounds (first 50 rounds with a 0.001 learning rate and then reduce the learning rate by 90\% in every tenth round) of training using the gradient solver ADAM. Similar to the previous scenario, we use the activation function ReLu in the input and the hidden layers. 
In the output layer, the linear activation function \footnote{Linear activation function: $\phi(x)=x$} is used for $m^2k^2$ neurons that correspond to the optimal Bell inequality and the sigmoid activation function is incorporated for the neuron that corresponds to the guessing probability. 
As the cost function, we use the Mean Squared Error (MSE) which is minimized during the training process.

\begin{figure}[ht]
\centering
\includegraphics[height=12cm, width=8cm]{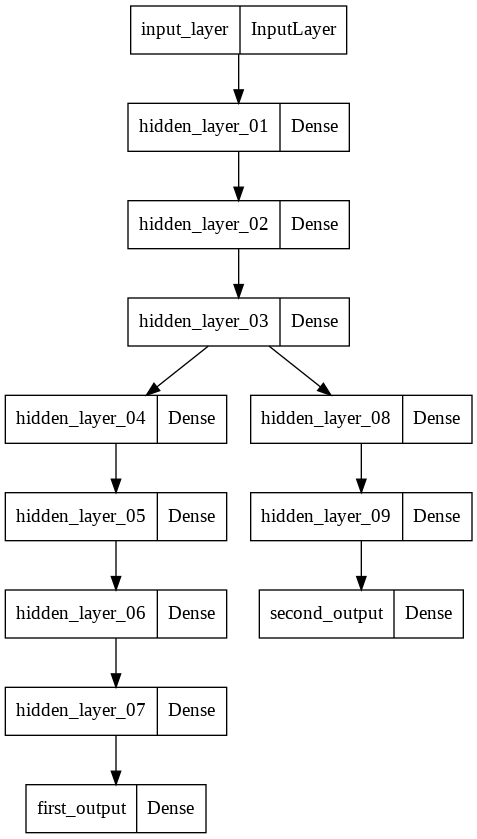}
\caption{Schematic diagram of a neural network where a hidden layer is bifurcated into two different arms which predict different parts of the output separately. In our scenario, input layer: $\{P(ab|xy)\}$ ($m^2k^2$ neurons), first output: $\{h_{abxy}\}$ ($m^2k^2$ neurons) and second output: $P^{*}_g(a|x,E)$ (1 neuron).}
\label{FFNN_twoNN}
\end{figure}

In addition, we use another neural network architecture with two parallel sub-models (by using branching) to interpret parts of the output that share the same input. In this construction, the input layer has $m^2k^2$ neurons corresponding to the elements of the probability distribution $\{P(ab|xy)\}$ of the $[m,k]$ Bell scenario. The input layer is followed by hidden layers consisting of multiple neurons. Then we bifurcate one hidden layer to create two branches. Several hidden layers then follow both branches; see Fig.~\ref{FFNN_twoNN} for visualization. The first branch of the network is for predicting the coefficients of the optimal Bell inequality $ \{h_{abxy}\}^{a,b=1,\cdots,k}_{x,y=1,\cdots,m} $ and thus has $m^2k^2$ neurons. The second branch of the network is for predicting the guessing probability. Thus, the output layer will have only one neuron corresponding to $P^{*}_g(a|x,E)$. In this paper, we refer to this neural network as $\mathrm{NN}_2$ which is built using the Keras functional API \cite{chollet2015}.    
In $\mathrm{NN}_2$, we use the ReLu activation function in the input and all the hidden layers. The linear activation function is used in the output layer of the first branch (which predicts the coefficients of the Bell inequality) while the sigmoid activation function is used in the second branch (which predicts the guessing probability). The other details of the training steps are the same as for the $\mathrm{NN}_1$ neural network stated previously. Both $\mathrm{NN}_1$ and $\mathrm{NN}_2$ predict the Bell inequality $B^{\text{pred}}$ (specified by the predicted coefficients $\{h_{abxy}^{\text{pred}}\}^{a,b=1,\cdots,k}_{x,y=1,\cdots,m}$) and the guessing probability $P^{\text{pred}}_g(a|x,E)$.

Since the neural networks $\mathrm{NN}_1$ and $\mathrm{NN}_2$ predict two separate entities (the optimal Bell inequality and the guessing probability), we evaluate their performance separately. 
We use the mean absolute error (see Eq.~(\ref{Eq. MAE})) and mean squared error (see Eq.~(\ref{Eq. MSE})) as our performance measure of predicting the guessing probability. The errors for different bipartite Bell scenarios are listed in Table~\ref{Table_probBIpGuess_pGuess_error}.    
\begin{table}[H]
\centering
\begin{tabular}{|c| c | c| c | c |} 
    \hline
    ANN & Metrics & [2,2] & [3,2] & [4,2]\\
    \hline
    \multirow{2}{*}{$\mathrm{NN}_1$} & MSE & 8.2$\times 10^{-6}$ & 0.002 & 0.009\\
    & MAE &  0.001  &  0.02 & 0.07\\ 	
    \hline
    \multirow{2}{*}{$\mathrm{NN}_2$} & MSE & 1.9$\times 10^{-7}$ & 0.001 & 0.002\\
    & MAE &  0.0003  &  0.013 & 0.027\\ 
    \hline 
\end{tabular}
\caption{Statistical errors of the predicted guessing probability $P^{\text{pred}}_g(a|x,E)$ with respect to the guessing probability $P^{*}_g(a|x,E)$ for different Bell scenarios for $\mathrm{NN}_1$ and $\mathrm{NN}_2$. Here the neural network is trained for predicting the guessing probability $P^{*}_g(a|x,E)$ and the Bell inequality $B$ from the probability distribution $P(ab|xy)$.}
\label{Table_probBIpGuess_pGuess_error}
\end{table}
Note that, for estimating the guessing probability, $\mathrm{NN}_2$ yields lower statistical errors than $\mathrm{NN}_1$. The reason lies in the structure of the neural network architectures. Since we create a branch in the neural network only to estimate the guessing probability, the $\mathrm{NN}_2$ neural network assigns more nodes to only estimate the guessing probability than $\mathrm{NN}_1$. 
In the case of predicting the optimal Bell inequality $B$ (characterized by its coefficients $\{h_{abxy}\}^{a,b=1,\cdots,k}_{x,y=1,\cdots,m}$), we use the performance measure MSE, which reads:
\begin{equation}
\begin{aligned}
&\text{MSE}\left[B,B^{\text{pred}}\right] \\
&= \frac{1}{m^2k^2} \frac{1}{N_{\mathrm{test}}}\sum_{i=1}^{N_{\mathrm{test}}}\sum_{a,b=1}^{k}\sum_{x,y=1}^{m} \left( \left(h_{abxy}\right)_i-(h_{abxy}^{\text{pred}})_i \right)^2 \, ,    
\end{aligned}
\end{equation}
and MAE, which reads:
\begin{equation}
\begin{aligned}
&\text{MAE}\left[B,B^{\text{pred}}\right] \\
&= \frac{1}{m^2k^2} \frac{1}{N_{\mathrm{test}}}\sum_{i=1}^{N_{\mathrm{test}}}\sum_{a,b=1}^{k}\sum_{x,y=1}^{m} \left| (h_{abxy})_i-(h_{abxy}^{\text{pred}})_i \right | \, .    
\end{aligned}
\end{equation}
The errors are listed in Table~\ref{Table_probBIpGuess_hyperplane_error}. 
\begin{table}[h]
\centering
\begin{tabular}{|c| c | c| c | c |} 
    \hline
    ANN & Metrics & [2,2] & [3,2] & [4,2]\\
    \hline
    \multirow{2}{*}{$\mathrm{NN}_1$} & MSE & 0.0004 & 0.0007 & 0.014\\
    & MAE &  0.001  &  0.002 & 0.067\\ 	
    \hline
    \multirow{2}{*}{$\mathrm{NN}_2$} & MSE & 0.0003 & 0.0005 & 0.015\\
    & MAE &  0.0009  &  0.002 & 0.069\\ 
    \hline 
\end{tabular}
\caption{Statistical errors of the coefficients of the predicted Bell inequality $B^{\text{pred}}$ (predicted by the trained deep learning models $\mathrm{NN}_1$ and $\mathrm{NN}_2$), $\{h_{abxy}^{\text{pred}}\}^{a,b=1,\cdots,k}_{x,y=1,\cdots,m}$ with respect to the original coefficients $\{h_{abxy}\}^{a,b=1,\cdots,k}_{x,y=1,\cdots,m}$ for different Bell scenarios.}
\label{Table_probBIpGuess_hyperplane_error}
\end{table}

Another way to evaluate the quality of the predicted Bell inequality is to use it for upper bounding the guessing probability problem (see Eq.~(\ref{SDP})). First, we estimate the probability of $P^{*}_g(a|x,E) < 1$, where $P^{*}_g(a|x,E)$ is calculated from the predicted Bell inequality $B^{\text{pred}}$ and the input-output probability distribution $\{P(ab|xy)\}$ of the test set. We present the results in Table~\ref{Table_hyperplane_pGuess_prob}. 	
\begin{table}[h]
\centering
\begin{tabular}{| c | c | c | c | c |} 
    \hline
    ANN & [2,2] & [3,2] & [4,2] \\
    \hline
    $\mathrm{NN}_1$ & 99.5\% & 98.7\% & 93.4\% \\
    \hline
    $\mathrm{NN}_2$ & 99.6\% & 99.4\% &  94.6\% \\
    \hline 
\end{tabular}
\caption{Probability of $P_g(a|x,E) < 1$ when using the Bell inequality $B^{\text{pred}}$ (predicted by the trained deep learning models $\mathrm{NN}_1$ and $\mathrm{NN}_2$).}
\label{Table_hyperplane_pGuess_prob}
\end{table}
We also look into the statistical errors between the original guessing probability $P^{*}_g(a|x,E)$ from the test set and the guessing probability calculated from the predicted Bell inequality $B^{\text{pred}}$. We use MAE and MSE as the performance measures listed in Table~\ref{Table_BIpredictedpGuess_pGuess_error}.   
\begin{table}[h]
\centering
\begin{tabular}{| c | c | c | c | c | c |} 
    \hline
    ANN & Metrics & [2,2] & [3,2] & [4,2] \\
    \hline
    \multirow{2}{*}{$\mathrm{NN}_1$} & MSE & 1.7 $\times 10^{-8}$ & 0.002 &  0.006 \\
    & MAE &  6.3 $\times 10^{-5}$  &  0.014 & 0.038 \\ 	
    \hline
    \multirow{2}{*}{$\mathrm{NN}_2$} & MSE & 1.5 $\times 10^{-8}$ & 0.002 & 0.004 \\
    & MAE &  4.1 $\times 10^{-5}$  &  0.014 & 0.031\\
    \hline 
\end{tabular}
\caption{Statistical errors between the guessing probability calculated from the predicted Bell inequality and the original guessing probability from the test set for various Bell scenarios and neural network constructions.}
\label{Table_BIpredictedpGuess_pGuess_error}
\end{table}
The high probability of generating $P^{*}_g(a|x,E) < 1$ with the predicted Bell inequalities (see Table~\ref{Table_hyperplane_pGuess_prob}) and the small statistical errors (see Table~\ref{Table_BIpredictedpGuess_pGuess_error}) demonstrate the quality and accuracy of the predicted Bell inequality.

We again compare the computational runtime of predicting the optimal Bell inequality using the standard linear optimization of Eq.~(\ref{linear_optimization}) with the neural network $\mathrm{NN}_1$ and $\mathrm{NN}_2$. The runtime for different methods is shown in Table~\ref{runtime comparison 2}.  
\begin{table}[h]
\centering
\begin{tabular}{|c|c|c|c|}\hline
    \diagbox[width=10em]{Time\\ per sample}{Bell \\ Scenario}&[2,2] & [3,2] & [4,2] \\ \hline
    LP & 95 ms & 159 ms & 379 ms \\ \hline
    $\mathrm{NN}_1$ & 29 $\mu$s  & 40 $\mu$s & 57 $\mu$s \\ \hline
    $\mathrm{NN}_2$ & 31 $\mu$s & 41 $\mu$s & 55 $\mu$s \\
    \hline
\end{tabular}
\caption{Runtime per sample comparison for estimation of the optimal Bell inequality using linear programming of Eq.~(\ref{linear_optimization}) and the neural network method for different Bell scenarios. LP stands for linear programming which is performed using the Mosek solver.}
\label{runtime comparison 2}
\end{table}
Similar to the previous scenario, the runtimes are evaluated over 10000 unknown samples and performed on the same personal computer in the same condition. 
The linear programming of Eq.~(\ref{linear_optimization}) is performed with the Mosek solver using PICOS \cite{sagnol2022picos} python interface.
We notice a significant speed-up when using the trained neural network models compared to the Mosek solver. 
This again follows from the fact that the number of variables in the optimization process of Eq.~(\ref{linear_optimization}) increases with the number of measurement settings (or outcomes per measurement) in the Bell scenario while the computational time for the neural networks only depends on its size.

\section{Discussion \& Conclusion}
Estimating the guessing probability is a cornerstone for device-independent quantum key distribution and device-independent randomness generation. This paper introduces a novel method to estimate the guessing probability using trained deep learning models to bypass the computationally complex and cumbersome semi-definite optimization process. 
Computation with the trained deep learning models is significantly faster than using a conventional solver. 
With current technology, Bell test event rates are around 100 kHz, which results in new data every 10$\mu$s \cite{bierhorst2018experimentally}. This frequency is too high for conventional SDP solvers on a single CPU. For those cases, our deep learning approach improves the computation significantly. In principle, optimizing the size of a deep neural network that can process each event as the experiment is being conducted is possible.

The deep learning model only provides an estimation of the upper bound of the guessing probability. But it will not provide a certification.    
Thus, additionally, our DL model provides an estimation of the optimal Bell inequality for which the Bell violation using the measurement statistics certifies the nonlocality of input-output correlations and guarantees that the guessing probability will be less than one.  
Our trained deep learning models, which significantly speed up the prediction of the Bell inequality compared to a conventional linear program solver, predict a Bell inequality that can generate $P_g(a|x,E) < 1$ with a very high probability. The mean average error between the guessing probability calculated from the predicted Bell inequality and the optimal Bell inequality (calculated using Eq.~(\ref{linear_optimization})) is in the order of $10^{-5}-10^{-2}$ (mean squared error is in the order of $10^{-8}-10^{-3}$) which shows the quality of this approach such that it can efficiently be used in a DIQKD or DIRNG protocol.

We also demonstrate a method for sampling random quantum correlations (correlations which have a realization of NPA hierarchy level of 2) using the facet Bell inequalities, which is then used as input in the supervised machine learning process. 
Note that, while generating probability distributions, we consider all facet Bell inequalities for the $[2,2]$ and $[3,2]$ Bell scenario. However, since there are more than 10000 facet Bell inequalities for the $[4,2]$ Bell scenario, we only restrict ourselves to generating probability distributions using the independent facet Bell inequalities.

To illustrate the benefits of our method, we have applied it to several relevant Bell scenarios.
Note that we design and train our neural networks to minimize statistical errors. However, we do not claim that our choice of the trained neural network is optimal for estimating the guessing probability and the associated optimal Bell inequality from the measurement statistics. Other constructions of neural networks will lead to different results. 
	
We observed that the statistical errors in the estimation of the guessing probability and the optimal Bell inequality increase with the complexity of the Bell scenario (i.e., the increase in the number of measurements per party). Since there are more inputs and outputs, our neural network architecture might not be able to generalize the extensive system with a limited number of hidden layers and nodes in each layer. To decrease the errors, one can take two steps. First, one can generate a larger dataset to train the model. Second, one can build a more extensive neural network architecture (i.e., more hidden layers or nodes in every layer). However, using a larger dataset for training or/and training a more extensive neural network will result in significantly more computational time. There is also the possibility of overfitting in an extensive network. A larger neural network architecture will also take more time to predict new instances. Therefore, one has to change the network architecture to optimize the speed and precision of a specific scenario.        
	
Note that while comparing the runtime for the Mosek optimization solver with the trained neural network for the estimation of the guessing probability (see Table~\ref{runtime comparison 1}), we implement the NPA hierarchy of level 2. The difference in computational runtime between the methods will be much more pronounced with increasing hierarchy.       
	
Our research demonstrates the applicability of deep learning techniques for Bell nonlocality and upper bounding the guessing probability. We believe that this strategy will create several research lines. The logical next step is to apply our approach to Bell scenarios with a higher number of measurement settings and outcomes. It is also possible to expand our framework to a multipartite scenario. Another direction worth exploring for future work is investigating other neural network constructions. Beyond the advantage in speed, one could use neural network architectures to search for new Bell inequalities. Also, recall that our methodology does not account for uncertainty or offers certification of the output. It remains for future work to use techniques like probabilistic modeling \cite{ghahramani2015probabilistic} that can certify the correctness of the model’s output.

\section{Acknowledgement}
The authors acknowledge support from the Federal Ministry of Education and Research BMBF (Project Q.Link.X). We thank Lucas Tendick for helpful discussions.

\bibliographystyle{ieeetr}
\bibliography{references}

\begin{thebibliography}{10}

\bibitem{bell1964einstein}
J.~S. Bell, ``$\text{On the Einstein Podolsky Rosen paradox}$,'' {\em Physics
  Physique Fizika}, vol.~1, no.~3, p.~195, 1964.

\bibitem{acin2007device}
A.~Ac{\'\i}n, N.~Brunner, N.~Gisin, S.~Massar, S.~Pironio, and V.~Scarani,
  ``Device-independent security of quantum cryptography against collective
  attacks,'' {\em Physical Review Letters}, vol.~98, no.~23, p.~230501, 2007.

\bibitem{pironio2009device}
S.~Pironio, A.~Acin, N.~Brunner, N.~Gisin, S.~Massar, and V.~Scarani,
  ``Device-independent quantum key distribution secure against collective
  attacks,'' {\em New Journal of Physics}, vol.~11, no.~4, p.~045021, 2009.

\bibitem{arnon2019simple}
R.~Arnon-Friedman, R.~Renner, and T.~Vidick, ``Simple and tight
  device-independent security proofs,'' {\em SIAM Journal on Computing},
  vol.~48, no.~1, pp.~181--225, 2019.

\bibitem{arnon2018practical}
R.~Arnon-Friedman, F.~Dupuis, O.~Fawzi, R.~Renner, and T.~Vidick, ``Practical
  device-independent quantum cryptography via entropy accumulation,'' {\em
  Nature Communications}, vol.~9, no.~1, pp.~1--11, 2018.

\bibitem{barrett2005no}
J.~Barrett, L.~Hardy, and A.~Kent, ``No signaling and quantum key
  distribution,'' {\em Physical Review Letters}, vol.~95, no.~1, p.~010503,
  2005.

\bibitem{masanes2011secure}
L.~Masanes, S.~Pironio, and A.~Ac{\'\i}n, ``Secure device-independent quantum
  key distribution with causally independent measurement devices,'' {\em Nature
  Communications}, vol.~2, p.~238, 2011.

\bibitem{vazirani2014fully}
U.~Vazirani and T.~Vidick, ``Fully device-independent quantum key
  distribution,'' {\em Physical Review Letters}, vol.~113, no.~14, p.~140501,
  2014.

\bibitem{masanes2014full}
L.~Masanes, R.~Renner, M.~Christandl, A.~Winter, and J.~Barrett, ``Full
  security of quantum key distribution from no-signaling constraints,'' {\em
  IEEE Transactions on Information Theory}, vol.~60, no.~8, pp.~4973--4986,
  2014.

\bibitem{acin2006efficient}
A.~Acin, S.~Massar, and S.~Pironio, ``Efficient quantum key distribution secure
  against no-signalling eavesdroppers,'' {\em New Journal of Physics}, vol.~8,
  no.~8, p.~126, 2006.

\bibitem{murta2019towards}
G.~Murta, S.~B. van Dam, J.~Ribeiro, R.~Hanson, and S.~Wehner, ``Towards a
  realization of device-independent quantum key distribution,'' {\em Quantum
  Science and Technology}, vol.~4, no.~3, p.~035011, 2019.

\bibitem{holz2020genuine}
T.~Holz, H.~Kampermann, and D.~Bru{\ss}, ``Genuine multipartite bell inequality
  for device-independent conference key agreement,'' {\em Physical Review
  Research}, vol.~2, no.~2, p.~023251, 2020.

\bibitem{hanggi2010device}
E.~H{\"a}nggi and R.~Renner, ``Device-independent quantum key distribution with
  commuting measurements,'' {\em arXiv preprint arXiv:1009.1833}, 2010.

\bibitem{pironio2010random}
S.~Pironio, A.~Ac{\'\i}n, S.~Massar, A.~B. de~La~Giroday, D.~N. Matsukevich,
  P.~Maunz, S.~Olmschenk, D.~Hayes, L.~Luo, T.~A. Manning, {\em et~al.},
  ``Random numbers certified by bell’s theorem,'' {\em Nature}, vol.~464,
  no.~7291, pp.~1021--1024, 2010.

\bibitem{nieto2018device}
O.~Nieto-Silleras, C.~Bamps, J.~Silman, and S.~Pironio, ``Device-independent
  randomness generation from several bell estimators,'' {\em New Journal of
  Physics}, vol.~20, no.~2, p.~023049, 2018.

\bibitem{pironio2013security}
S.~Pironio and S.~Massar, ``Security of practical private randomness
  generation,'' {\em Physical Review A}, vol.~87, no.~1, p.~012336, 2013.

\bibitem{bancal2014more}
J.-D. Bancal, L.~Sheridan, and V.~Scarani, ``More randomness from the same
  data,'' {\em New Journal of Physics}, vol.~16, no.~3, p.~033011, 2014.

\bibitem{nieto2014using}
O.~Nieto-Silleras, S.~Pironio, and J.~Silman, ``Using complete measurement
  statistics for optimal device-independent randomness evaluation,'' {\em New
  Journal of Physics}, vol.~16, no.~1, p.~013035, 2014.

\bibitem{bischof2017measurement}
F.~Bischof, H.~Kampermann, and D.~Bru{\ss}, ``Measurement-device-independent
  randomness generation with arbitrary quantum states,'' {\em Physical Review
  A}, vol.~95, no.~6, p.~062305, 2017.

\bibitem{acin2012randomness}
A.~Ac{\'\i}n, S.~Massar, and S.~Pironio, ``Randomness versus nonlocality and
  entanglement,'' {\em Physical Review Letters}, vol.~108, no.~10, p.~100402,
  2012.

\bibitem{acin2016certified}
A.~Ac{\'\i}n and L.~Masanes, ``Certified randomness in quantum physics,'' {\em
  Nature}, vol.~540, no.~7632, pp.~213--219, 2016.

\bibitem{skrzypczyk2018maximal}
P.~Skrzypczyk and D.~Cavalcanti, ``Maximal randomness generation from steering
  inequality violations using qudits,'' {\em Physical Review Letters},
  vol.~120, no.~26, p.~260401, 2018.

\bibitem{navascues2007bounding}
M.~Navascu{\'e}s, S.~Pironio, and A.~Ac{\'\i}n, ``Bounding the set of quantum
  correlations,'' {\em Physical Review Letters}, vol.~98, no.~1, p.~010401,
  2007.

\bibitem{navascues2008convergent}
M.~Navascu{\'e}s, S.~Pironio, and A.~Ac{\'\i}n, ``A convergent hierarchy of
  semidefinite programs characterizing the set of quantum correlations,'' {\em
  New Journal of Physics}, vol.~10, no.~7, p.~073013, 2008.

\bibitem{carrasquilla2017machine}
J.~Carrasquilla and R.~G. Melko, ``Machine learning phases of matter,'' {\em
  Nature Physics}, vol.~13, no.~5, pp.~431--434, 2017.

\bibitem{broecker2017machine}
P.~Broecker, J.~Carrasquilla, R.~G. Melko, and S.~Trebst, ``Machine learning
  quantum phases of matter beyond the fermion sign problem,'' {\em Scientific
  Reports}, vol.~7, no.~1, pp.~1--10, 2017.

\bibitem{canabarro2019machine}
A.~Canabarro, S.~Brito, and R.~Chaves, ``Machine learning nonlocal
  correlations,'' {\em Physical Review Letters}, vol.~122, no.~20, p.~200401,
  2019.

\bibitem{carleo2017solving}
G.~Carleo and M.~Troyer, ``Solving the quantum many-body problem with
  artificial neural networks,'' {\em Science}, vol.~355, no.~6325,
  pp.~602--606, 2017.

\bibitem{deng2018machine}
D.-L. Deng, ``Machine learning detection of bell nonlocality in quantum
  many-body systems,'' {\em Physical Review Letters}, vol.~120, no.~24,
  p.~240402, 2018.

\bibitem{gao2017efficient}
X.~Gao and L.-M. Duan, ``Efficient representation of quantum many-body states
  with deep neural networks,'' {\em Nature Communications}, vol.~8, no.~1,
  pp.~1--6, 2017.

\bibitem{ma2018transforming}
Y.-C. Ma and M.-H. Yung, ``Transforming bell’s inequalities into state
  classifiers with machine learning,'' {\em NPJ Quantum Information}, vol.~4,
  no.~1, pp.~1--10, 2018.

\bibitem{mehta2019high}
P.~Mehta, M.~Bukov, C.-H. Wang, A.~G. Day, C.~Richardson, C.~K. Fisher, and
  D.~J. Schwab, ``A high-bias, low-variance introduction to machine learning
  for physicists,'' {\em Physics Reports}, 2019.

\bibitem{torlai2018neural}
G.~Torlai, G.~Mazzola, J.~Carrasquilla, M.~Troyer, R.~Melko, and G.~Carleo,
  ``Neural-network quantum state tomography,'' {\em Nature Physics}, vol.~14,
  no.~5, pp.~447--450, 2018.

\bibitem{venderley2018machine}
J.~Venderley, V.~Khemani, and E.-A. Kim, ``Machine learning out-of-equilibrium
  phases of matter,'' {\em Physical Review Letters}, vol.~120, no.~25,
  p.~257204, 2018.

\bibitem{datta2021device}
S.~Datta, H.~Kampermann, and D.~Bru\ss{}, ``Device-independent secret key rates
  via a postselected bell inequality,'' {\em Phys. Rev. A}, vol.~105,
  p.~032451, Mar 2022.

\bibitem{brunner2014bell}
N.~Brunner, D.~Cavalcanti, S.~Pironio, V.~Scarani, and S.~Wehner, ``Bell
  nonlocality,'' {\em Reviews of Modern Physics}, vol.~86, no.~2, p.~419, 2014.

\bibitem{pitowsky1982resolution}
I.~Pitowsky, ``Resolution of the einstein-podolsky-rosen and \text{Bell}
  paradoxes,'' {\em Physical Review Letters}, vol.~48, no.~19, p.~1299, 1982.

\bibitem{pitowsky1991correlation}
I.~Pitowsky, ``Correlation polytopes: their geometry and complexity,'' {\em
  Mathematical Programming}, vol.~50, no.~1-3, pp.~395--414, 1991.

\bibitem{Lofberg2004}
J.~L{\"{o}}fberg, ``Yalmip: A toolbox for modeling and optimization in
  matlab,'' in {\em In Proceedings of the CACSD Conference}, (Taipei, Taiwan),
  2004.

\bibitem{gb08}
M.~Grant and S.~Boyd, ``Graph implementations for nonsmooth convex programs,''
  in {\em Recent Advances in Learning and Control} (V.~Blondel, S.~Boyd, and
  H.~Kimura, eds.), Lecture Notes in Control and Information Sciences,
  pp.~95--110, Springer-Verlag Limited, 2008.
\newblock \url{http://stanford.edu/~boyd/graph_dcp.html}.

\bibitem{cvx}
M.~Grant and S.~Boyd, ``{CVX}: Matlab software for disciplined convex
  programming, version 2.1.'' \url{http://cvxr.com/cvx}, Mar. 2014.

\bibitem{qetlab}
N.~Johnston, ``{QETLAB}: A {MATLAB} toolbox for quantum entanglement, version
  0.9.'' \url{http://qetlab.com}, Jan. 2016.

\bibitem{Goodfellow-et-al-2016}
I.~Goodfellow, Y.~Bengio, and A.~Courville, {\em Deep Learning}.
\newblock MIT Press, 2016.
\newblock \url{http://www.deeplearningbook.org}.

\bibitem{krivachy2021high}
T.~Kriv{\'a}chy, Y.~Cai, J.~Bowles, D.~Cavalcanti, and N.~Brunner, ``High-speed
  batch processing of semidefinite programs with feedforward neural networks,''
  {\em New Journal of Physics}, vol.~23, no.~10, p.~103034, 2021.

\bibitem{clauser1969proposed}
J.~F. Clauser, M.~A. Horne, A.~Shimony, and R.~A. Holt, ``Proposed experiment
  to test local hidden-variable theories,'' {\em Physical Review Letters},
  vol.~23, no.~15, p.~880, 1969.

\bibitem{fukuda2003cddlib}
K.~Fukuda, ``Cddlib reference manual,'' {\em Report version 093a, McGill
  University, Montr{\'e}al, Quebec, Canada}, 2003.

\bibitem{collins2004relevant}
D.~Collins and N.~Gisin, ``A relevant two qubit bell inequality inequivalent to
  the chsh inequality,'' {\em Journal of Physics A: Mathematical and General},
  vol.~37, no.~5, p.~1775, 2004.

\bibitem{pal2010maximal}
K.~F. P{\'a}l and T.~V{\'e}rtesi, ``Maximal violation of a bipartite
  three-setting, two-outcome bell inequality using infinite-dimensional quantum
  systems,'' {\em Physical Review A}, vol.~82, no.~2, p.~022116, 2010.

\bibitem{cruzeiro2019complete}
E.~Z. Cruzeiro and N.~Gisin, ``Complete list of tight bell inequalities for two
  parties with four binary settings,'' {\em Physical Review A}, vol.~99, no.~2,
  p.~022104, 2019.

\bibitem{cope2019bell}
T.~Cope and R.~Colbeck, ``Bell inequalities from no-signaling distributions,''
  {\em Phys. Rev. A}, vol.~100, p.~022114, Aug 2019.

\bibitem{barrett2005nonlocal}
J.~Barrett, N.~Linden, S.~Massar, S.~Pironio, S.~Popescu, and D.~Roberts,
  ``Nonlocal correlations as an information-theoretic resource,'' {\em Physical
  review A}, vol.~71, no.~2, p.~022101, 2005.

\bibitem{popescu1994quantum}
S.~Popescu and D.~Rohrlich, ``Quantum nonlocality as an axiom,'' {\em
  Foundations of Physics}, vol.~24, no.~3, pp.~379--385, 1994.

\bibitem{kingma2014adam}
D.~P. Kingma and J.~Ba, ``Adam: A method for stochastic optimization,'' {\em
  arXiv preprint arXiv:1412.6980}, 2014.

\bibitem{aps2019mosek}
M.~ApS, ``Mosek optimization toolbox for matlab,'' {\em User’s Guide and
  Reference Manual, Version}, vol.~4, 2019.

\bibitem{wittek2015algorithm}
P.~Wittek, ``Algorithm 950: Ncpol2sdpa—sparse semidefinite programming
  relaxations for polynomial optimization problems of noncommuting variables,''
  {\em ACM Transactions on Mathematical Software (TOMS)}, vol.~41, no.~3,
  pp.~1--12, 2015.

\bibitem{chollet2015}
F.~Chollet, ``keras.'' \url{https://github.com/fchollet/keras}, 2015.

\bibitem{sagnol2022picos}
G.~Sagnol and M.~Stahlberg, ``Picos: A python interface to conic optimization
  solvers,'' {\em Journal of Open Source Software}, vol.~7, no.~70, p.~3915,
  2022.

\bibitem{bierhorst2018experimentally}
P.~Bierhorst, E.~Knill, S.~Glancy, Y.~Zhang, A.~Mink, S.~Jordan, A.~Rommal,
  Y.-K. Liu, B.~Christensen, S.~W. Nam, {\em et~al.}, ``Experimentally
  generated randomness certified by the impossibility of superluminal
  signals,'' {\em Nature}, vol.~556, no.~7700, pp.~223--226, 2018.

\bibitem{ghahramani2015probabilistic}
Z.~Ghahramani, ``Probabilistic machine learning and artificial intelligence,''
  {\em Nature}, vol.~521, no.~7553, pp.~452--459, 2015.

\end{thebibliography}

\end{document}